\documentclass{article}
\usepackage[table,xcdraw]{xcolor}
\usepackage{spconf,amsmath,graphicx}
\usepackage{amsmath}
\usepackage{amssymb}
\usepackage{url}
\usepackage{array}
\usepackage{amssymb}
\usepackage{multirow}
\usepackage{graphicx}
\usepackage{tabularx}

\title{FLAP: Fast Language-Audio Pre-training}
\name{
Ching-Feng Yeh, Po-Yao Huang, Vasu Sharma, Shang-Wen Li and Gargi Gosh
}
\address{
\texttt{\{cfyeh,berniehuang,vasusharma,shangwel,gghosh\}@meta.com}
\\
FAIR, Meta}

\begin{document}

\maketitle

\begin{abstract}
We propose Fast Language-Audio Pre-training (FLAP), a  self-supervised approach that efficiently and effectively learns aligned audio and language representations through masking, contrastive learning and reconstruction.
For efficiency, FLAP randomly drops audio spectrogram tokens, focusing solely on the remaining ones for self-supervision. 
Through inter-modal contrastive learning, FLAP learns to align paired audio and text representations in a shared latent space.
Notably, FLAP leverages multiple augmented views via masking for inter-modal contrast and learns to reconstruct the masked portion of audio tokens.
Moreover, FLAP leverages large language models (LLMs) to augment the text inputs, contributing to improved performance.
These approaches lead to more robust and informative audio-text representations, enabling FLAP to achieve state-of-the-art (SoTA) performance on audio-text retrieval tasks on \textit{AudioCaps} (achieving 53.0\% R@1) and \textit{Clotho} (achieving 25.5\% R@1).
\end{abstract}


\section{Introduction}
\label{sec:intro}

Representation learning~\cite{representation} has garnered significant momentum on creating information-rich embeddings for downstream tasks. 
Recently, self-supervised representation learning (SSL)~\cite{ssl,NIPS2014_ssl} has emerged as a prominent research area in hope of reducing human annotations.
Traditionally, SSL approaches have been developed under the single-modality setup for image~\cite{mocov3, mae}, text~\cite{bert,roberta}, or audio/speech~\cite{amae, hsu2021hubert, jaiswal2020survey} independently.
However, there is a growing interest in representation learning across multiple modalities~\cite{clip, videoclip, mmp, stica}, which brings both challenges and exciting new possibilities. 
One breakthrough is Contrastive Language-Image Pre-training (CLIP) \cite{clip} which projects text and image embeddings into a shared latent space, enabling applications like cross-modality retrieval and automatic captioning.
More recently, Contrastive Language-Audio Pre-training (CLAP) \cite{elizalde2023clap, laionclap2023} learns representations for both text and audio and delivered strong performance on audio-text retrieval tasks. 

The key ingredients in CLIP and CLAP are their SSL objectives and model architectures.
On objective, both CLIP and CLAP utilize contrastive learning, which aims to minimize the distance between embeddings in different modalities of the same instance, while differentiating the embeddings from different instances \cite{invariant_mapping, simclr, he2020momentum}. 
On model architecture, both CLIP and CLAP adopted Transformer-like models \cite{transformer}, which have proven to be effective.
Previous studies suggest this transformer$+$contrastive learning combination produces high-quality embeddings for both uni-modal ~\cite{mocov3, simclr, wav2vec2} and multi-modal~\cite{stica, mavil, wav2clip} tasks. 
One major limitation of Transformer-like models is their quadratic complexity with respect to sequence lengths, which becomes a computational bottleneck and restricts overall efficiency.

To improve computational efficiency, techniques with Masked AutoEncoders (MAE) such as image MAE~\cite{mae}, VideoMAE \cite{tong2022videomae, feichtenhofer2022masked} and AudioMAE~\cite{amae} were recently proposed and achieved significant efficiency wins with minor performance trade-off.
Recently, Fast Language-Image Pre-training (FLIP) \cite{li2023scaling} applied similar techniques to image-text SSL.
Recognizing that audio signals in nature are continuous and with variable in lengths, we explored the masking strategies for self-supervised language-audio representation learning. 
We term our model Fast Language-Audio Pre-training (FLAP).
FLAP endeavors to establish aligned audio and language representations by incorporating masking, contrastive learning and reconstruction techniques.
For language-audio datasets, very often the audio signals contain much richer information than the text counterparts.
For example, an audio segment of dog barking may reveal additional information such as volume and frequency, which are often missing in the text. Also, text descriptions can vary in writing styles and generate inconsistent embeddings for the same semantics. Given such imbalanced information richness between audio and text, we utilize large language models (LLMs) \cite{touvron2023llama, vicuna2023, alpaca} to enrich and unify the writing style for texts in the language-audio task. 

Previous works \cite{laionclap2023, deshmukh2022audio, Koepke2022, Mei2022metric} on language-audio pre-training received wide research interests. Recently, large-scale CLAP (LS-CLAP) \cite{laionclap2023} demonstrated strong results on audio-text retrieval on \textit{AudioCaps} and \textit{Clotho} benchmarks. 
In this study, we further improve the LS-CLAP results by 1) using Masked Audio-Video Learners (MAViL) as pre-trained audio encoder 2) efficient masking for efficiency and robustness 3) adding audio reconstruction for better embedding 4) utilizing LLMs for text augmentation.
We observed significant performance boosts from FLAP, which outperformed the recently proposed state-of-the-art systems \cite{laionclap2023}.

\begin{figure*}[t]
    \centering
    \includegraphics[scale=0.5]{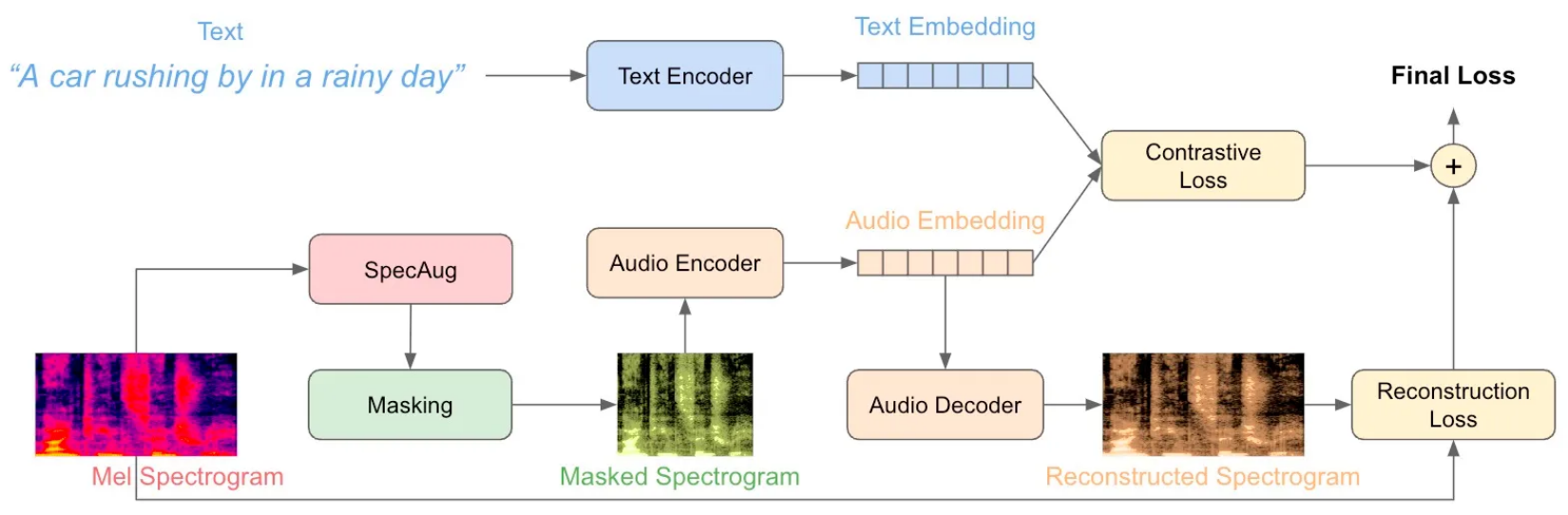}
    \caption{The architecture of FLAP, including audio/text encoders, efficient masking and audio reconstruction.}
    \label{fig:audioclip-arch}
    \vspace{-0.5cm}
\end{figure*}

\section{Contrastive Learning}
\label{sec:contrastive}

The fundamental framework of contrastive learning involves selecting an ``anchor'' data sample, a data point from the same distribution referred to as the ``positive'' sample, and a data point from a different distribution known as the ``negative'' sample. Contrastive learning aims to reduce the distance between the anchor and positive samples, which are part of the same distribution, in the latent space. Simultaneously, it seeks to maximize the distance between the anchor and the negative samples. 
For learning aligned audio and text representations, the ``positive'' examples refers to the representations of paired audio and text samples (i.e., an audio and its corresponding captions), while the negative examples are all the combinations of the unpaired audios and captions sampled in a batch. 
In this work we employ the InfoNCE \cite{oord2018representation} loss for inter-modal contrastive learning over audio and text pairs sampled from a dataset $(\mathbf{a}, \mathbf{t})\in\mathcal{D}$.
Let $\mathbf{a}$ and $\mathbf{t}$ respectively denote the instance-level audio and text representations. The InfoNCE loss $\mathcal{L}_{\text{c}}(\mathbf{a},\mathbf{t})$ is defined as:
\begin{equation}
\label{eq:contrastive}
\mathcal{L}_{\text{c}}(\mathbf{a},\mathbf{t}) = 
-\frac{1}{B} \sum_{i=1}^B {\rm log}  
\frac{ {\rm exp} (\text{S}(\mathbf{a}_i,\mathbf{t}_i)/\tau)}
{\sum_{j=1}^{B} {\rm exp} (\text{S}(\mathbf{a}_i,\mathbf{t}_j)/\tau)) },
\end{equation}
where $\text{S}(\mathbf{a}_i,\mathbf{t}_j)=\frac{\mathbf{a}_i^T\mathbf{t}_j}{\|\mathbf{a}_i\| \|\mathbf{t}_j\|}$ is the cosine similarity between $\mathbf{a}_i$, $\mathbf{t}_j$ and $\tau$ is the softmax temperature. In Eq \ref{eq:contrastive}, the loss function encourages the distance between the embeddings from audio and text from the same sample to be minimized and to be maximized from different samples, therefore achieving the desired "contrasting" effects. It is worth noting that the performance of contrastive learning depends highly on the number of samples ($B$) being contrasted against each other within the same batch. Larger batch sizes ($B$) offers more inter-sample connections to stabilize the aggregated gradients for updating model parameters, with increased need of computation and memory consumption.

\section{FLAP: Efficient Masking}
\label{sec:masking}

Inspired by the recent success of FLIP\cite{li2023scaling} which attempts to employ the masking technique for learning image-text representations, we propose Fast Language-Audio Pre-training (FLAP) for learning self-supervised audio-language representations by employing masking for both contrastive learning and reconstruction. 
As depicted in Fig.~\ref{fig:audioclip-arch}, FLAP consists of an audio encoder, a text encoder, and audio decoder.
For FLAP's audio encoder, we adopt the audio backbone from MAViL \cite{mavil}, the SoTA audio model pre-trained on the audio and video clips of AudioSet~\cite{audioset}.
MAViL is a self-supervised framework for learning audio-video representations that comprises two stages. 
In the first stage, MAViL simultaneously learns to reconstruct spectrogram and pixels, leveraging the complementary information from both modalities. 
In the second stage, MAViL performs self-distillation where a student model predicts the contextualized features generated by the first-stage teacher model. 

FLAP performs instance-wise inter-modal contrastive learning using Eq.~\ref{eq:contrastive} over the non-masked (visible) portion of audio spectrogram tokens. The masking strategy in FLAP significantly enhances the computation efficiency and promotes more robust representation as masking can also be viewed as a data augmentation approach over the audio tokens. 
Specifically, given a input tensor of shape $(B, N, D)$, where $B$ is the batch size, $N$ is the sequence length, $D$ is the embedding dimension, masking reduces the shape to $(B, N', D)$, where $N'$ is smaller than $N$.
This enables significant computation reduction for Transformer-like models as the model complexity grows quadratically with sequence length (i.e. $O(N^2)$). 

\begin{figure}[t]
    \centering
    \includegraphics[scale=0.27]{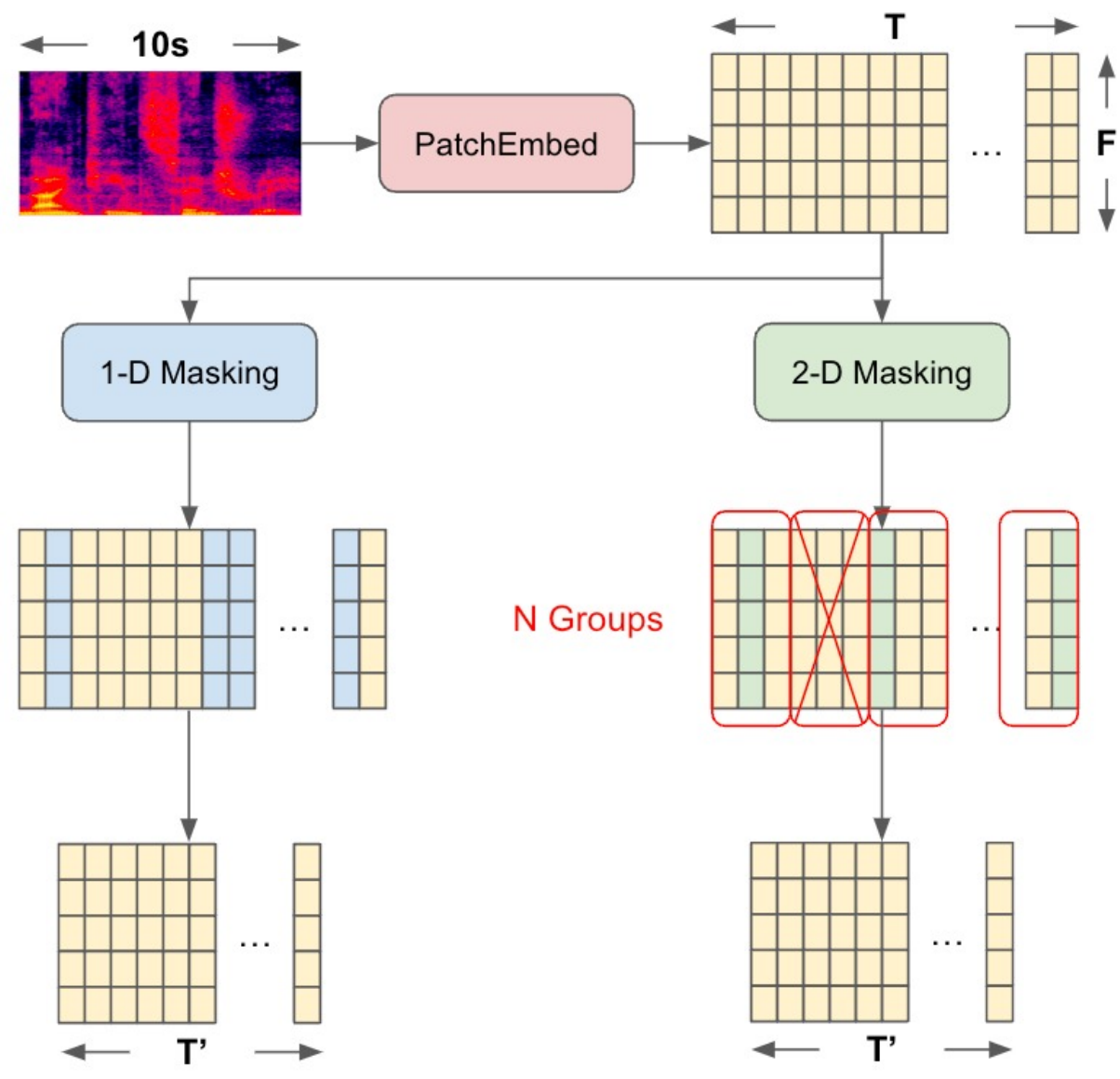}
    \caption{Frame dropping by 1-D and 2-D Masking.}
    \label{fig:masking}
    \vspace{-0.5cm}
\end{figure}

We investigated two masking strategies, namely 1-D and 2-D masking, as illustrated in Fig. \ref{fig:masking}.
Before masking, the input (in the form of mel-spectrogram) is transformed into patch embeddings.
For 1-D masking, the input tensor of shape $(B, N, D)$ is first augmented with positional embeddings and then randomly sampled on the T-axis to become $(B, N', D)$.
The random sampling is performed on a shuffled and per-frame basis to the desired length $N'$.
1-D masking is simple and effective in boosting robustness by random frame dropping and reducing computation along with sequence lengths $N$.
On the other hand, 2-D masking aims to build a more structured sampling strategy on top of 1-D masking.
Instead of directly sampling on the $N$-axis, 2-D masking first splits the $N$-axis into $M$ groups, each having $K = N/M$ consecutive frames.
Next, both the $M$ groups and the $K$ frames in each group are sampled individually in the same fashion as in 1-D masking and reduced to $M'$ and $K'$ respectively.
Finally, both $M'$ and $K'$ are merged back together and becomes the new $N' = M' * K'$.
2-D masking essentially splits the overall sequence ($N$) into numerous ($M$) fine-grained segments ($K$), therefore enables more structured sampling through both homogeneous sampling and dropping in each fine-grained segments.
Both 2-D and 1-D maskings can achieve similar efficiency improvement with different masking ratios.
For example, a 75\% masking ratio on $N$ leads to 25\% (= 100\% - 75\%) computation cost for 1-D masking, while 50\% on $M$ and 50\% on $K$ for 2-D masking also leads to 25\% (= 50\% * 50\%). 
The masked tensors are then directly sent to the audio encoder for computing the output embeddings for each frame and then averaged across the $N$-axis for per-instance embeddings.
These masking strategies are particularly useful for contrastive learning tasks as the per-example outputs are more robust to frame dropping. 
In addition, reduced sequence length by masking also enables larger batch sizes to fit in GPUs, which benefits contrastive learning as more pairs are involved in the loss function for a single batch.
Furthermore, the masking strategy can be view as a type of audio augmentation (e.g. SpecAug~\cite{park2019specaugment}) that promotes robustness of the learned representations.
The masking is applied during the training stage and is disabled during evaluation. 

\begin{table*}[t]
\footnotesize
\begin{tabularx}{\textwidth}{|p{.2\linewidth}|p{.2\linewidth}|X|}
\hline
Original & LLM-only & AED+LLM \\ \hline
Wind blows as waves crash against a shoreline. 
& The wind gusts while the waves crash against the shoreline 
& The waves crash against the shoreline, with the sound of wind blowing and creating wind noise. The ocean is in motion, with the wind blowing strongly \\\hline
A man is speaking. 
& A person is delivering a speech. 
& A man's speech can be heard rustling the leaves in the wind.
\\\hline
A loud siren whizzes past. 
& A piercing siren blares by.
& An ambulance (siren) with a loud siren sound whizzes pasts by.\\\hline
An engine revving. 
& A car's engine revving. 
& An accelerating car engine revving with a vroom sound.\\\hline
Water sound.
& The sound of water.
& An intermittent sound of water flowing from a tap or faucet.\\\hline
The sound of a boat.
&
A boat is making sound.
&
A water vehicle, specifically a motorboat or speedboat, is moving at a rapid pace with winds blowing.\\\hline
\end{tabularx}
\caption{Comparison between original captions and augmented captions generated by LLM and AED.}\label{tab:llm}

\end{table*}

\section{Audio Reconstruction}
\label{sec:mlm}

To bolster the robustness of the learned audio embeddings, we further propose an additional objective that promotes the incorporation of audio information into the embeddings. 
This can be achieved by tasking the model with reconstructing the original audio spectrogram tokens using the per-sample embeddings.
As depicted in Fig. \ref{fig:audioclip-arch}, before being aggregated across sequence length to produce the per-sample audio embeddings, the per-frame embeddings (of shape $(B, T', D)$) is sent to an audio decoder for reconstructing the mel-spectrogram. 
Empirically, we observe that reconstructing only the spectrogram but \textit{not} the text tokens yields better performance.
We employ vanilla Transformer blocks as the audio $f_\text{a}^{-1}(.)$ decoders. 
The encoder's outputs ($\mathbf{a}_{\text{mm}}$ are firstly projected and padded with trainable \texttt{[MASK]} tokens.
After restoring the original order (time-frequency for audio and space-time for video tokens), we add the decoders' (fixed 2-D sinusoidal) positional embeddings and input the restored sequences into the decoders.
At the top of the decoders, we incorporate linear heads to reconstruct the raw inputs.
Specifically, the decoder outputs for spectrogram reconstruction are denoted as $\mathbf{\hat{a}}=f_{\text{a}}^{-1}(g_{\text{av}}(f_{\text{a}}(\mathbf{a}')))$. For notation clarity, we omit the \texttt{[MASK]} tokens and linear projection head.
Let $\hat{\mathbf{a}}_i,\mathbf{a}_i^{\text{raw}} \in \mathbb{R}^{H_{\text{raw}}^{\text{a}}}; i=1\dots n$ 
denote the audio decoder's output and the ground truth reference of the $i$-th masked spectrogram patch. 
In masked audio reconstruction, FLAP is self-supervised by minimizing the mean squared error (MSE) loss $\mathcal{L}_r^{\text{raw}}$ defined as:
\begin{equation}
    \mathcal{L}_r^{\text{raw}} = \frac{1}{n}\sum_{i=1}^n(\hat{\mathbf{a}}_i-\mathbf{a}_i^{\text{raw}})^2 
\end{equation}

The MSE loss from reconstruction is then weighted and added to the final loss along with the contrastive loss. 
With reconstruction, the model is encouraged to preserve condensed information into per-sample embeddings, as these embeddings not only have to be close to their text domain counterparts, but also useful in producing original inputs. 
It is worth noting that reconstruction does come with a trade-off on efficiency and batch size, as the audio decoder requires non-trivial computation and memory usage. 

\section{Enriched Augmentation by LLM}
\label{sec:llm}

Learning audio-text representations faces an additional challenge stemming from the scarcity of audio-text pairs in existing audio-text corpora.
Collecting human annotations for audio is both expensive and non-scalable.
To address this issue, we present a novel approach that harnesses the power of large language models (LLMs) and audio event detection models (AEDs) to augment the limited number of text descriptions available for audio.
Table~\ref{tab:llm} shows examples of the original text descriptions from training data and the class list to caption transformation. 
From the original text descriptions, it is clear that both the richness of information is behind the corresponding audio signals and the writing styles are inconsistent across samples.
To reinterpret and enrich the same semantic for natural language, we leverage the power of LLMs~\cite{touvron2023llama,vicuna,alpaca} to enhance the descriptiveness of the audio captions on audio-text datasets such as \textit{AudioCaps} and \textit{Clotho}, which only contains weak and limited descriptive captions. 
We first employ off-the-shelf AED model (i.e., MAViL~\cite{mavil}) to detect the audio events within a sample. And then we exploits a LLM (i.e., Vicuna~\cite{vicuna}) along with engineered prompts to combine the classification outputs and the original caption to generate richer captions for samples in \textit{AudioCaps} and \textit{Clotho}.
Vicuna is an open-source instruction-following model fine-tuned on the Llama-7b model~\cite{llama}. From the examples in Table \ref{tab:llm}, utilizing this model generates more grammatical captions that remain faithful to the audio events.

To enrich text captions with LLM and detected audio events, we used the following prompt: ``Describe a situation with \textit{AED results} sounds and combine it with the \textit{original\_caption} together." A limitation of the Vicuna model is its tendency to add unnecessary details or ignore relevant labels when generating captions. By adding AED outputs and original captions into the prompt, we leveraged the in-context learning ability of Vicuna to enrich captions.
During training, the same set of audio signals with text descriptions replaced with generated captions are augmented to the datasets. 

\begin{table*}[t]
\vspace{-0.2cm}
\resizebox{\textwidth}{!}{
\begin{tabular}{lcccccccc|cccccc}
\hline\hline
\multicolumn{1}{c}{\multirow{3}{*}{\textbf{Model}}} &
  \multirow{3}{*}{\textbf{Global Batch Size}} &
  \multirow{3}{*}{\textbf{Masking}} &
  \multicolumn{6}{c|}{\textbf{AudioCaps  Eval.}} &
  \multicolumn{6}{c}{\textbf{Clotho Eval.}} \\ \cline{4-15} 
\multicolumn{1}{c}{} &
   &
   &
  \multicolumn{3}{c}{\textbf{T-A Retrieval}} &
  \multicolumn{3}{c|}{\textbf{A-T Retrieval}} &
  \multicolumn{3}{c}{\textbf{T-A Retrieval}} &
  \multicolumn{3}{c}{\textbf{A-T Retrieval}} \\ \cline{4-15} 
\multicolumn{1}{c}{}  &  &  & R@1 & R@5 & R@10 & R@1 & R@5  & R@10 & R@1  & R@5  & R@10 & R@1  & R@5  & R@10 \\ \hline
(1) LS-CLAP\cite{laionclap2023}    & 2304 & -- &  32.7    &  68.0  &  81.2   &  43.9   &  77.7    &  87.6    &  15.6    &   38.6  &  52.3    & 23.7    &  48.9   &  59.9    \\ \hline
(2) FLAP & 2304 (36 x 64 GPUs) & -- & 34.8 & 70.0 & 82.7 & 49.0 & 79.5 & 88.7 & 16.3 & 41.4 & 53.9 & 23.0 & 49.2 & 61.4 \\
(3) FLAP & 2304 (36 x 64 GPUs) & 1-D: 0.4 & 36.0 & 70.5 & 83.0 & 49.0 & 78.9 & 89.2 & 16.8 & 40.7 & 53.4 & 23.9 & 48.9 & 61.2 \\
(4) FLAP (+recon) & 2304 (36 x 64 GPUs) & 1-D: 0.4 & 36.7 & 71.2 & 83.3 & 47.2 & 81.9 & 90.0 & 15.6 & 39.5 & 51.9 & 21.7 & \textbf{50.6} & 61.3 \\
(5) FLAP & 2304 (36 x 64 GPUs) & 2-D: 0.2/0.2 & 37.5 & 73.5 & 84.6 & 49.6 & 82.3 & 89.4 & 17.2 & 41.1 & 52.8 & \textbf{23.7} & 48.7 & 62.3 \\
(6) FLAP (+recon) & 2304 (36 x 64 GPUs) & 2-D: 0.2/0.2 & 37.2 & 73.0 & 84.9 & 50.3 & 81.4 & 90.0 & 17.0 & 41.2 & 53.5 & 22.4 & 49.0 & 62.7 \\
(7) FLAP & 4608 (72 x 64 GPUs) & 2-D: 0.2/0.2 & \textbf{38.3} & \textbf{73.6} & \textbf{85.1} & \textbf{50.6} & \textbf{83.1} & \textbf{91.2} & \textbf{16.7} & \textbf{41.5} & \textbf{54.2} & 23.0 & 48.6 & \textbf{62.9} \\ \hline\hline
\end{tabular}}
\vspace{-0.2cm}
\caption{Experimental results on masking type, masking ratio and audio reconstruction (without feature fusion).}
\label{tab:mask_and_recon}
\end{table*}

\section{Experiments}
\label{sec:exp}

\subsection{Datasets and Setup}
\label{ssec:setup}

Across all experiments, similar to LS-CLAP \cite{laionclap2023}, we use AudioCaps, Clotho, and 5 other datasets (Freesound, Epidemic Sound, BBC Sound Effects, Free To Use Sounds, Sonniss Game effects) for training, while AudioCaps \cite{audiocaps} and Clotho \cite{clotho} are used for evaluation. 
It is worth noting that compared to LS-CLAP, we drop AudioStock due to its unavailability and therefore the size of the dataset for training is smaller than LS-CLAP. 
The evaluation sets are identical for fair comparisons.
We built experiments on top of the LS-CLAP \cite{laionclap2023} toolkit and adopted fvcore \cite{fvcore} for efficiency analysis. 
Cross-modality retrieval between audio and text is used for evaluation of the quality of the embeddings. 
For text-audio (T-A) retrieval, given the text as query, the audio recordings in the evaluation set are ranked based on the cosine similarities between text and audio embeddings.
The same procedure applies to audio-text (A-T) retrieval.
Recalls at top 1, 5 and 10 (R@1, R@5 and R@10) are reported as metrics for both tasks on \textit{AudioCaps} and \textit{Clotho} datasets. 

For experiments without feature fusion, depending on the audio length, we either randomly chunk 10 seconds from longer audios or pad to 10 seconds for shorter ones to form input data of uniform lengths.
For feature extraction, 25ms window size and 10ms window shift were used to extract mel-spectrogram features with 128 mel-bins.
For experiments with feature fusion enabled, we followed the same procedure as LS-CLAP \cite{laionclap2023}, where audios are either padded or strided to create global and local versions followed by 2-D convolutions for merging.
For SpecAug \cite{park2019specaugment}, up to 192 audio frames (e.g. 1.92 seconds) and up to 48 mel-bins are randomly replaced with zeros for each sample.
For text embedding generation, the texts paired with audio data are tokenized with a capped length of 77.
RoBERTa \cite{roberta} is used as text encoder for all experiments to be consistent with LS-CLAP \cite{laionclap2023}. 

The Adam \cite{kingma2014adam} optimizer with $\beta_1=0.99$, $\beta_2=0.9$ was used during model training.
The learning rate starts with a warm-up stage, peaks at $10^{-4}$ and was decayed on a cosine schedule until the target number of epochs (45) is reached.
Since both masking or reconstruction affects GPU memory usage which translates to largest batch size allowed per GPU, we report results with similar batch sizes to the baseline (2304) and also results with larger batch sizes enabled by efficient masking but using the equivalent computational resources (i.e. the same number of GPUs). 

\subsection{Results on Efficient Masking and Reconstruction}
\label{ssec:retrieval}

To evaluate the performance of efficient masking and reconstruction, the experimental results are summarized in Table \ref{tab:mask_and_recon}, in which all results are without feature fusion.
The results from LS-CLAP \cite{laionclap2023} are listed in row 1 serving as the baseline.
In row 2, the audio encoder is replaced with the recent MAViL \cite{mavil} model with audio-and-video self-supervised pre-training that achieves state-of-the-art performance on audio classification tasks.
Note that we simply train MAViL with the contrastive loss Eq. \ref{eq:contrastive} on audio-text datasets without masking or reconstruction applied.
The results validate that stronger audio-modal yields improved audio-text representations in audio-text retrieval tasks.
In rows 3 and 5, 1-D and 2-D masking are applied with masking ratio selected from additional ablation studies, 0.4 for 1-D and 0.2/0.2 for 2-D respectively.
For 2-D masking, we split the sequence into 64 (e.g. $N$ = 64) groups of 8 (e.g. $K$ = 8) frames from patch embeddings of length 256.
From the comparison, we observed similar sequence length reduction from 1-D (1 - 0.4 = 60\%) and 2-D ((1 - 0.2) $\times$ (1 - 0.2) = 64\%).
But 2-D masking delivers better improvement due to more structured masking strategy.
Both 1-D and 2-D masking reduce memory usage and preserve room for additional operations.
On top of masking, audio reconstruction is applied with 4 layers of Transformer decoding layers with 4 heads and 512 embedding dimensions for each layer.
The results with audio reconstruction are listed in rows 4 and 6.
The reconstruction objective encourages FLAP to capture more abstract concepts from audio context to represent and predict raw audio spectrograms, without replying on additional class labels.
This results in stronger audio-text retrieval performance on \textit{AudioCaps}.
Alternatively, the memory saving from masking can be also utilized to process more samples in a single batch instead of audio reconstruction.
Doubling the batch size produces the results in row 7. 
Compared with rows 5 and 6, increasing the batch size improves the robustness of the contrastive objective.
In Eq.\ref{eq:contrastive}, the positive pairs are encouraged to contrast against a larger collection of negative samples in the denominator, resulting in more well-aligned audio-text latent space where semantically correlated audio-text pairs are closer to each other and uncorrelated ones are distant.
For contrastive learning, sufficiently large batch size is crucial to the model performance.
It is worth noting that the number of GPUs are kept the same across the comparisons and larger batch size is achieved through efficient masking, which not only improves the robustness of the model but also reduces computation and memory footprints. 

\subsection{Efficiency Analysis of 1-D/2-D Masking}
\label{ssec:efficiency}

\begin{figure}[t]
    \centering
    \includegraphics[scale=0.25]{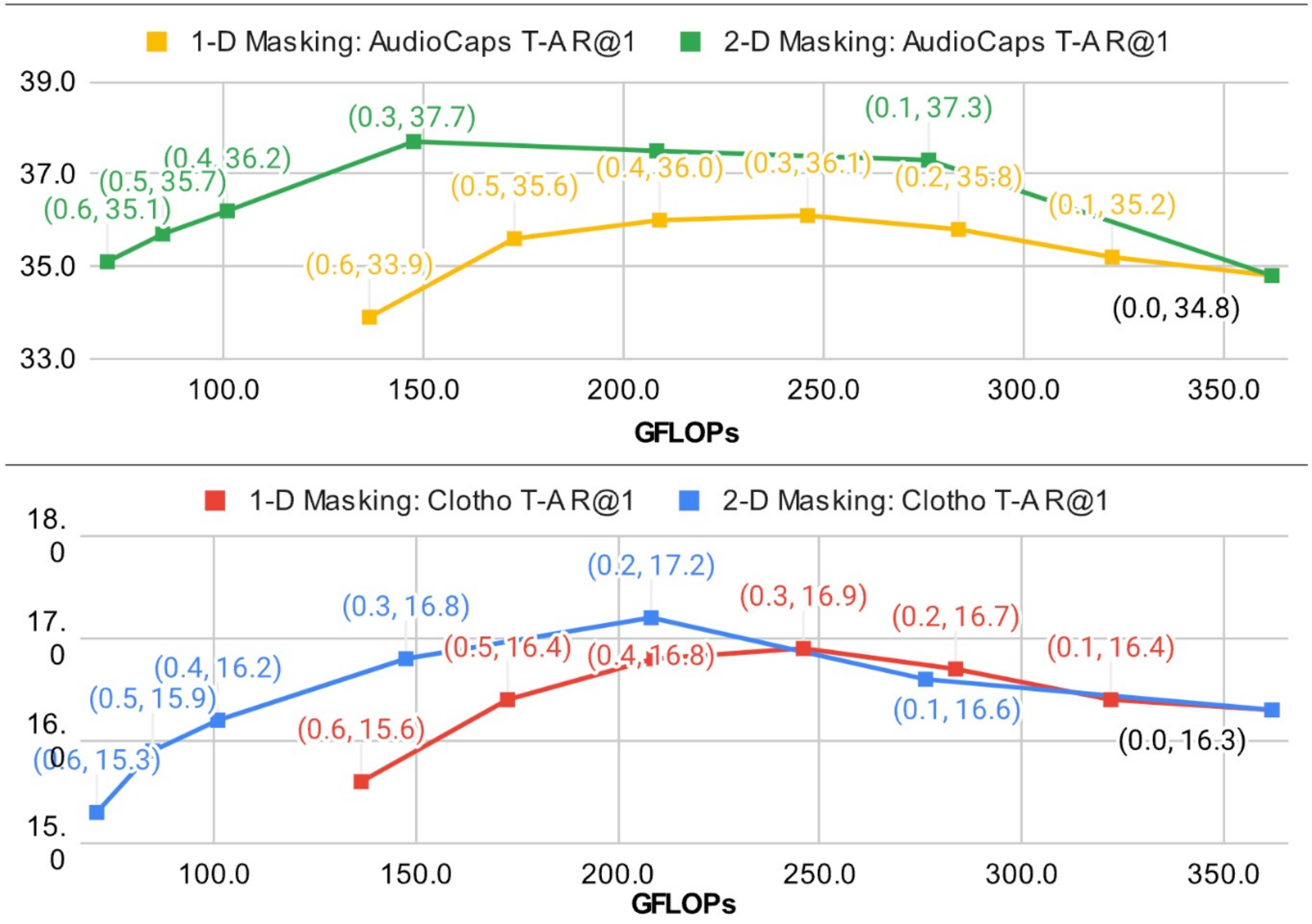}    
    \caption{Text-Audio R@1 vs. GFLOPs on \textit{AudioCaps} and \textit{Clotho} with Different Ratios for 1-D and 2-D Masking.}
    \label{fig:efficiency}
    \vspace{-0.5cm}
\end{figure}

Masking provides benefits including bringing down the sequence length for efficiency and improvement on model robustness.
However, similar to many efficiency-focused approaches, the typical efficiency/performance tradeoff also applies here.
To analyze the correlation between masking ratios and the impact on model performance, models with different masking strategies and incremental masking ratios are trained and compared in operational curves in Fig. \ref{fig:efficiency}, with \textit{AudioCaps} results on top and \textit{Clotho} results at bottom.
In the operational curves, the computation complexity (in terms of GFLOPs) serves as the horizontal axis while the top 1 recall in retrieval (in terms of R@1) serves as the vertical axis.
We also annotate each data point with \textit{(masking ratio, R@1)} for easier numerical comparison.
The GFLOPs are calculated using the fvcore \cite{fvcore} tool for the audio encoder only with a batch of 8 samples of 10-second lengths. The batch size was kept the same for all masking ratios for fair comparison.
The baseline results with no masking are also included at the rightmost positions in Fig. \ref{fig:efficiency}.

For each dataset, 1-D and 2-D masking are compared with incremental masking ratios.
The masking ratios are on a per-dimension basis, meaning for the same masking ratio, 2-D masking presents more aggressive frame dropping.
For example, when masking ratio is 0.3, 1-D masking preserves 70\% of the sequence while 2-D masking only preserves 49\% (= 0.7 * 0.7).
From the curves in Fig. \ref{fig:efficiency}, efficient masking started improving model robustness until too many frames in the sequence are dropped (around 0.5).
This is expected as information loss is increased along with the masking ratio.
In addition, similar to the observation in Table \ref{tab:mask_and_recon}, 2-D masking provides better recalls around similar GFLOPs therefore offers better trade-off than 1-D masking for being more structured.
Taking masking ration = 0.2 for example, 2-D masking approximately saves 25\% of the computation and delivers better recalls than result without masking.
This shows that the efficient masking is effective in both improving efficiency and model robustness.

\begin{table*}[t]
\vspace{-0.2cm}
\resizebox{\textwidth}{!}{
\begin{tabular}{lcccccccc|cccccc}
\hline\hline
\multicolumn{1}{c}{\multirow{3}{*}{\textbf{Model}}} &
  \multirow{3}{*}{\shortstack[c]{\textbf{Feature} \\ \textbf{Fusion}}} &
  \multirow{3}{*}{\shortstack[c]{\textbf{Batch} \\ \textbf{Size}}} &
  \multicolumn{6}{c|}{\textbf{AudioCaps  Eval.}} &
  \multicolumn{6}{c}{\textbf{Clotho Eval.}} \\ \cline{4-15} 
\multicolumn{1}{c}{} &
   &
   &
  \multicolumn{3}{c}{\textbf{T-A Retrieval}} &
  \multicolumn{3}{c|}{\textbf{A-T Retrieval}} &
  \multicolumn{3}{c}{\textbf{T-A Retrieval}} &
  \multicolumn{3}{c}{\textbf{A-T Retrieval}} \\ \cline{4-15} 
\multicolumn{1}{c}{}  & & & R@1   & R@5  & R@10 & R@1  & R@5  & R@10 & R@1  & R@5  & R@10 & R@1  & R@5  & R@10 \\ \hline
(1) LS-CLAP\cite{laionclap2023} & \multirow{5}{*}{No} & 2304 &  32.7    &  68.0  &  81.2   &  43.9   &  77.7    &  87.6    &  15.6    &   38.6  &  52.3    & 23.7    &  48.9   &  59.9    \\ \cline{4-15} 
(2) FLAP & & 2304 & 37.5 & 73.5 & 84.6 & 49.6 & 82.3 & 89.4 & 17.2 & 41.1 & 52.8 & 23.7 & 48.7 & 62.3 \\
(3) FLAP (+recon) & & 2304 & 37.2 & 73.0 & 84.9 & 50.3 & 81.4 & 90.0 & 17.0 & 41.2 & 53.5 & 22.4 & 49.0 & 62.7 \\
(4) FLAP & & 4608 & 38.3 & 73.6 & \textbf{85.1} & 50.6 & \textbf{83.1} & 91.2 & 16.7 & \textbf{41.5} & \textbf{54.2} & \textbf{23.0} & 48.6 & 62.9 \\ 
(5) FLAP (+LLM-aug) & & 4608 & \textbf{40.4} & \textbf{74.7} & 85.0 & \textbf{51.5} & 82.5 & \textbf{92.5} & \textbf{17.4} & 41.3 & 53.7 & 21.6 & \textbf{51.2} & \textbf{63.1} \\  \hline
(6) LS-CLAP\cite{laionclap2023} & \multirow{5}{*}{Yes} & 2304 & 36.2   &  70.3  &  82.5   & 45.0 & 76.7 & 88.0 & 17.2 & 42.9 & 55.4 & 24.2 & 51.1 & 66.9 \\ \cline{4-15} 
(7) FLAP & & 2304 &  38.6 & 74.2 & 85.6 & 49.6 & 83.8 & 91.1 & 17.3 & 43.1 & 55.7 & 24.4 & 53.2 & 66.4 \\
(8) FLAP (+recon) & & 2304 & 40.1 & 74.8 & 86.0 & 50.8 & 81.9 & 91.0 & 17.8 & 44.0 & 56.3 & 24.6 & 53.0 & 66.7 \\
(9) FLAP & & 4608 & 39.9 & 75.4 & \textbf{86.6} & 50.6 & 81.7 & 91.9 & 17.5 & 43.4 & 56.0 & 24.4 & 52.1 & 67.1 \\ 
(10) FLAP (+LLM-aug) & & 4608 & \textbf{41.5} & \textbf{75.5} & 86.0 & \textbf{53.0} & \textbf{84.1} & \textbf{92.6} & \textbf{20.3} & \textbf{46.5} & \textbf{58.8} & \textbf{25.5} & \textbf{53.4} & \textbf{67.9} \\  \hline\hline
\end{tabular}}
\vspace{-0.2cm}
\caption{Experimental Results on Feature Fusion and Text Augmentation with Large Language Models (LLM). \textsuperscript{*}NOTE: FLAP uses the same dataset as LS-CLAP ~\cite{laionclap2023}, excluding AudioStock due to its unavailability.}
\label{tab:fusion}
\end{table*}

\subsection{Results on Feature Fusion and LLM Augmentation}
\label{ssec:fusion_and_llm}

The setup without feature fusion in LS-CLAP adds padding for shorter audio signals and applies random cropping for longer ones to generate inputs to the model of uniform lengths of 10 seconds.
It works well for feeding long audio signals to the audio encoder without increasing the computational complexity.
However, random cropping also implies information loss.
Therefore, feature fusion \cite{laionclap2023} was introduced to further enhance the final retrieval performance and achieved significant improvements. 
To evaluate FLAP on the same setup, we adopted the same feature fusion and the corresponding results are listed in Table \ref{tab:fusion}. 
In Table \ref{tab:fusion}, results without feature fusion are listed in rows 1 to 5 and results with feature fusion are in rows 6 to 10.
Rows 1 to 5 share same setups in Table \ref{tab:mask_and_recon}, where row 1 is the same CLAP baseline, row 2 is the 2-D masked MAViL with ratio 0.2, row 3 incorporates reconstruction loss on top of row 2, row 4 doubles the batch size compared with row 2 and row 5 augments LLM-generated text descriptions on top of row 4.
Rows 6 to 10 repeats the same setups as rows 1 to 5 except inputs with feature fusion were used. 

Compared with rows 1 to 5, rows 6 to 10 are effectively improved with feature fusion, as feature fusion combines global and cropped segments as inputs to the model.
This benefits more for long audio signals, as observed from the larger improvements on \textit{Clotho}, which contains more audio segments longer than 10 seconds. 
Comparing rows 7-10 to row 6, FLAP delivers similar performance improvement for feature fusion setups similarly to rows 2-5.
This demonstrates that FLAP is highly versatile and adds complementary gains on top of the already competitive feature fusion results.   
In rows 5 and 10, LLM augmentation mentioned in section \ref{sec:llm} is also applied on top of the best models to demonstrate the impact from enriched and more consistent text descriptions.
Compared with rows 4 and 9, results with augmentation from LLM-generated text descriptions show either similar or better performance. 
Particularly, the results on \textit{Clotho} with feature fusion showed larger improvement.
Since the enriched text description tends to be longer as observed from examples in Table \ref{tab:llm}, feature fusion setups potentially benefit more for better audio-text match and alignment.
Rows 5 and 10 also serve as the best results with and without feature fusion for the proposed FLAP framework. 
Compared with CLAP, combining efficient masking which leads to increased batch sizes along with enriched text description by LLMs yields significant improvements across both text-audio and audio-text retrieval tasks on both datasets. 
For top 1 recall (R@1), FLAP in row 5 without feature fusion performs better on majority of tasks than the previous best results with feature fusion in row 6 (36.2 to 40.4 for text-audio and 45.0 to 51.5 for audio-text on \textit{AudioCaps}, 17.2 to 17.4 for text-audio on Clotho, with exception on 24.2 to 21.6 for audio-text on \textit{Clotho}).
On the same feature fusion setup, FLAP in row 10 further outperforms the previous best results in row 6 on all tasks (36.2 to 41.5 for text-audio and 45.0 to 53.0 for audio-text on \textit{AudioCaps}, 17.2 to 20.3 for text-audio and 24.2 to 25.5 for audio-text on \textit{Clotho}). 
To the best of our knowledge, these results also serve as the current best performances on audio-text and text-audio retrieval tasks for \textit{AudioCaps} and \textit{Clotho}.

\section{Conclusion}
\label{sec:conclusion}

In this paper, we introduce Fast Language-Audio Pre-training (FLAP) where contrastive learning meet masking.
FLAP leads to better audio understanding, task performance, and enables efficient and effective learning on sequence modalities such as audio and video.
In addition, audio reconstruction and enriched text description augmentation by large language models (LLMs) are also investigated.
Efficient masking reduces both computation and memory footprint for training samples, therefore enables larger batch sizes for contrastive learning.
Text augmentation from LLMs further enriches the text descriptions for audio signals and produces more consistent writing styles.
Combining both, FLAP delivers strong performance on audio-text retrieval tasks with evaluation on \textit{AudioCaps} and \textit{Clotho} benchmarks.
The techniques in FLAP are versatile and applicable to representation learning in sequence modalities such as text, audio and video.

\bibliographystyle{IEEEbib}
\bibliography{strings,refs}

\end{document}